\documentclass[12pts]{article} 
\usepackage[latin1]{inputenc}
\usepackage [dvips]{graphicx} 
\newcommand{\ee}{\end{equation}}
\newcommand{\be}{\begin{equation}}

\begin{document}
\title{A ``Little Big Bang'' Scenario of Multifragmentation}

\author{X. Campi, H. Krivine,\\{\small\it LPTMS\footnote{UMR 8626, CNRS
Universit\'e de Paris XI}, F-91405 Orsay Cedex, France}\\ E.
Plagnol\footnote{eric.plagnol@cdf.in2p3.fr} \\{\small\it Collège de
France/PCC, F-75231 Paris Cedex 05, France }\\and \\N.
Sator\footnote{Present adress : LPTL, Université de Paris VI, 4, place Jussieu 75252, Paris Cedex 05 France} \\{ \small\it Dipartimento di Fisica,
Universit\`a di Napoli "Federico II",} \\ {\small\it INFM Napoli, Via
Cintia, 80126 Napoli, Italy}}

\maketitle

\begin{abstract}

We suggest a multifragmentation scenario in which fragments are
produced at an early, high temperature and high density, stage of the
reaction. In this scenario, self-bound clusters of particles in the
hot and dense fluid are the precursors of the observed fragments. This
solves a number of recurrent problems concerning the kinetic
energies and the temperature of the fragments, encountered with the
standard low density fragmentation picture. The possibility to recover
the initial thermodynamic parameters ($T$ and $\rho$) from the
inspection of the asymptotic fragment size and kinetic energy
distributions is discussed.
\end{abstract}

\section{Introduction}

Recent theoretical studies of the morphology of simple fluids 
have suggested the presence of self-bound clusters
of particles \cite{campi2,campi1,sator1}. The (time averaged) mass
distributions of these clusters depend essentially on the energy and
density of the system and resemble those found in nuclear multifragmentation
($U$-shapes, power law, exponentials). Furthermore,  these
distributions (event averaged) remain invariant if the system is
allowed to expand freely.

The above observations suggest the following scenario for
multifragmentation: In a first step, the nucleus is excited (and possibly
compressed). In this "hot" and dense phase, at least part of it reaches
thermal equilibrium. Clusters, defined as self-bound ensembles of particles,
are present in this medium. Immediately after, the system starts to expand,
out of equilibrium, as \emph{an ensemble of interacting clusters}.
Asymptotically, these \emph{clusters} cease to interact with each other,
becoming the observed \emph{fragments}. The sudden expansion reveals the
cluster distribution of the primordial hot and dense phase of the system
(\emph{a Little Big Bang}) and ``freezes'' it. The long range Coulomb
repulsion between clusters helps this freezing process. The smallness of the
system also helps, because most of the clusters are close to the surface and
escape freely into the vacuum. The clusters have ramified shapes in the
dense medium and become, at asymptotic times, spherical fragments.

The "standard" mechanism \cite{MMC,SMM,Friedman,GSI_Ns} of
multifragmentation, adopted in the so-called "Statistical Equilibrium
Models" (SEM), is different: Once the system is excited, it undergoes an
homogeneous and quasi-static expansion. At some sufficiently low density
(freeze-out), it "recondensates" as \emph{an ensemble of spherical
non-overlapping and non-interacting fragments}. At this stage, the models
calculate the fragment mass distributions as those of a gas of
non-interacting fragments with internal nuclear structure, at equilibrium.
These models have been extremely successful in describing the observed
fragment mass distributions \cite{GSI_Ns,INDRA_Ns}. It is however important
to notice that the main hypothesis (equilibrium at low freeze-out density,
spherical fragments) required in these models to make the calculations
feasible, are not substantiated by other, more microscopic, approaches of
the multifragmentation phenomenon \cite{CMD1,CMD2,CMD3,QMD1,QMD2}. In fact,
despite the success in describing the mass distributions, this "standard"
scenario leads to some recurrent problems. Among them, the most important is
the prediction of the kinetic energies of the fragments that are too low
compared with experimental observations \cite{Ekins1,Ekins2}. This problem
is "corrected" by adding an \emph{ad hoc} radial flow to the fragments.
However the origin of this flow is unclear and seems inconsistent with the
hypothesis of equilibrium and the low freeze-out density. Another problem
concerns the internal temperature of the fragments, which in the experiments
seems to be much lower than the temperature of the gas of fragments
\cite{B_Tsang,Serfling}. We will see that the proposed high density scenario
with pre-existing clusters helps to solve these problems in a very natural
way.

This approach is substantiated by Classical Molecular Dynamics (CMD)
calculations \cite{campi2,campi1,CMD1,CMD2,CMD3} of an ensemble of particles
either confined in a container or freely expanding. Our goal is to interpret
the (well known) results obtained with these methods at the light of our
present knowledge on clusters in hot and dense fluids. We are thus not
proposing a well-finished model of nuclear multifragmentation which could be
directly compared with experimental data, but rather a generic scenario.

This paper is organized as follows. In section 2, we
 present results on the mapping between thermodynamics and clustering in
 dense and confined medium, in large and in nuclear-size
 systems. In  section 3 we discuss the free expansion of these
 latter systems. Some final remarks will be found  in Section 4.

\section{Clusters in Confined Systems}

In order to investigate the mapping between thermodynamics and
clustering, we first have to define what a physical cluster is, over the whole
range of densities and temperatures. A definition solely based
on  a proximity criterium (interaction radius) in configuration space is
obviously not adequate at high densities and/or temperatures.

At low density, a cluster is naturally defined as a self bound ensemble of
particles. By continuity we shall extend this definition to a dense medium.
Several approaches can be used to identify self-bound clusters among the
particles of the system. For instance, they can be defined as a partition of
particles which minimizes the interaction energy between clusters
\cite{randrup,PUE}. Clusters can also be built, bond by bond, using Hill's
criterium \cite{hill}: Two particles are bound if their relative kinetic
energy is less than their potential energy. In other words, a cluster is a
set of particles close to each other in phase space and \emph{not only} in
configuration space. It must be emphasized that, with this model, knowing
the positions and velocities of the particles, no adjusting parameter is
necessary to recognize the clusters whatever the density or the temperature.
In previous works \cite{campi2,campi1,sator1}, we checked that the cluster
size distribution does not depend significantly on the specific definition
of self-bound clusters mentioned above.

The mapping between thermodynamics and clustering can be studied by
performing classical microcanonical molecular dynamic simulations of a
simple fluid. Particles confined in a cubic container interact through a
Lennard-Jones potential ($V(r)=4\epsilon((\sigma/r)^{12}-(\sigma/r)^{6})$),
where $\epsilon$ and $\sigma$ define the energy and length scales
respectively. The natural time-unit is defined by
$\tau_0=\sqrt{m\sigma^2/(48\epsilon)}$, where  $m$
is the mass of the particles. When Coulomb interaction is
included, it is done with the prescription of reference \cite{CMD1}.
Self-bound clusters are recognized by means of Hill's criterium. The details
of the calculations can be found in references \cite{campi2,campi1,sator1}.

In order to establish an accurate mapping between thermal and
geometrical properties of this fluid, we first present results for the
case of a large system ($N\simeq 12000$ particles) with periodic
boundary conditions \cite{campi2} (and, of course without Coulomb interaction):
\begin{itemize}
\item When crossing the condensation curve at sub-critical densities, a
macroscopic cluster appears in the liquid-gas coexistence
region\footnote{This correspondence between thermodynamics and
clustering is not observed when clusters are simply defined by a
constant interaction radius in the configurational space.}. Although
defined in a complete different theoretical framework, our clusters
mark the liquid-gas coexistence line as Fisher's
clusters do \cite{fisher}\footnote{Notice however that Fisher's clusters
are non-interacting and not microscopically defined.}. 
\item A percolation line starts at the thermodynamical critical point
(within the uncertainties inherent to the critical slowing down) and goes
through the supercritical region of the phase diagram. Thus, a power law
behavior of the cluster size distribution does not necessarily imply that
the system is in its thermodynamical critical state. Also notice that this
percolation line does not extend below the critical temperature.
\item It can be shown that the Hill prescription we are using to define
self-bound clusters in a realistic fluid is equivalent \cite{campi5} to the
prescription introduced by Coniglio and Klein \cite{coniglio} to define
physical clusters in the lattice-gas model. This gives a theoretical
framework to understand the existence of this critical line. Along this line
(usually called Kertész line), the critical exponents are those of the
random percolation, except at the thermodynamical critical point, where the
exponents are those of the correlated percolation \cite{coniglio}.
\item The energy of the system is nearly constant along the critical
percolation line.
\end{itemize}
\begin{figure}[h!] 
\begin{center}
\includegraphics[angle=0,scale=0.3]{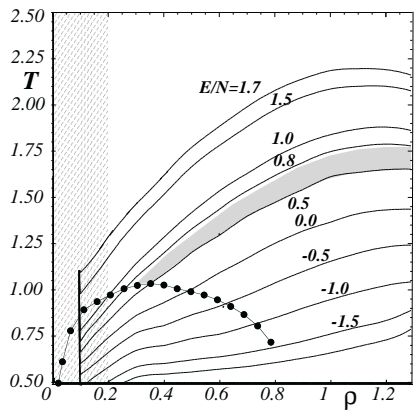}
\end{center}
\caption{\label{fig2} \it \footnotesize Phase diagram of a Lennard-Jones
fluid (including Coulomb interaction) of $N=189$ particles confined in a
container. The position of the coexistence line is sketched by filled dots.
Lines of equal energy are indexed by the energy per particle of the system.
The shaded area indicates the location of the percolation critical region. The
dashed area shows tentatively the domain of application of the standard SEM.
Temperature $T$ and density $\rho$ are in units of the Lennard-Jones
parameters ($\epsilon$ and $\sigma$). The full circle indicates the critical point. }
\end{figure}
These results remain valid in small systems \cite{campi1}.
We have studied the behaviour of small systems like atomic nuclei by
performing similar calculations with $N=189$ particles confined in a cubic
container with perfectly reflecting walls and interacting \emph{via} a Lennard-Jones potential plus Coulomb. The ``phase diagram'' is
represented in figure \ref{fig2}. Although the sharp effects found in the
thermodynamical limit are smoothed by the finite size and surface effects,
we reach basically the same conclusions.

\begin{figure}[h!] 
\begin{center}
\includegraphics[scale=.3,angle=-0]{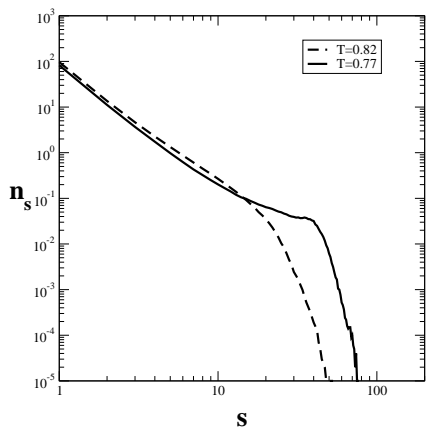}
\end{center}
\caption{\label{fig4} \it \footnotesize Cluster size distributions
$n_s$ just above ($T=0.82$) and below ($T=0.77$) the coexistence
``line'' at $\rho=0.1$ for a system of $N=189$ particles.}
\end{figure}

\begin{itemize}
\item As is shown in figure \ref{fig4}, the
shape of the cluster size distribution changes suddenly when crossing
the condensation curve for a very small variation of temperature. 
\item The cluster size distribution presents a power law behavior along a
percolation ``line'' (more precisely a band due to the finite size effects)
which starts at the thermodynamical critical point (region) (see figure
\ref{fig2}). Along this line, the slope of the cluster size distribution
($\tau\simeq 2.5$) is in accordance with percolation theory\footnote{Due to
the range of our potential, in a small system, all particles interact with
each other, so the mean field result is expected.}.
\end{itemize}
We note also that, in the supercritical region, a given cluster
size distribution corresponds qualitatively  to a given energy of the
system. We stress that the converse is not true, notably
below the coexistence line.

\begin{figure}
\vskip 0.5cm
\begin{center}
\includegraphics[angle=-0,scale=.35]{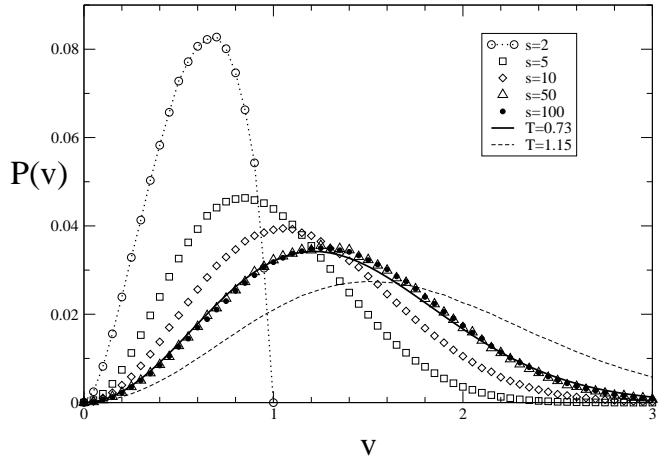} 
\end{center}
\caption{\label{fig55}\it Internal velocity distribution for different cluster
sizes $s$ (for $\rho=0.50$ and $T=1.15$). For the purposes of
comparison, the Maxwell distribution function is represented for
$T=1.15$ (temperature of the system) and for $T=0.73$ (asymptotic
internal effective temperature of the clusters, see figure \ref{fig5}).}
\end{figure}

\begin{figure}[h!] 
\begin{center}
\includegraphics[width=7.5cm,angle=-0]{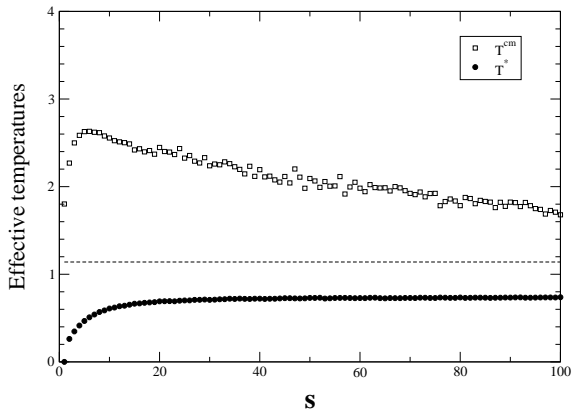}
\end{center}
\caption{\label{fig5} \it \footnotesize Effective internal
temperature of clusters $T^{*}(s)$ and temperature of the gas of clusters
$T_{cm}(s)$ as a function of the cluster size $s$, for
$\rho=0.50$ and $T=1.15$. The real temperature of the fluid (dashed
line) is indicated for comparison. The system contains $N=189$
particles.}
\vskip 0.7cm
\end{figure}

Studying the degrees of freedom associated to the
internal motion of the particles inside the clusters, and the degrees
of freedom associated to the center of mass motion of these clusters,
we define two "effective temperatures", $T^{*}(s)$ and $T^{cm}(s)$, as
two third of the average kinetic energy of the corresponding ensemble
\footnote{However, if we were able to isolate a cluster from the others
(this is what happens during the expansion), the internal effective
temperature would become the real thermodynamic temperature of the isolated
fragment.}. Let us define the internal velocity $v^{*}$ as the velocity of a
particle calculated in the center of mass system of the cluster it belongs
to. As is shown in Fig \ref{fig55}, the internal velocity distribution for
cluster of size $s=2$ contrasts strongly with a Maxwell distribution. As it
should be, by definition of the self-bound clusters, the distribution is
truncated at $v^{*}=1$. However, increasing the cluster size, the velocity
distribution tends to a Maxwell distribution characterized by the asymptotic
temperature $T^{*}(s)$ for $s>50$ (see figure \ref{fig5}). Similar results
\cite{SOT98} have been recently obtained by calculating the first order
corrections (namely the interaction between a monomer and a self-bound
cluster) to a perfect gas of clusters at low density. Whatever the size of
the clusters and the thermodynamic state of the system, the internal
effective temperature $T^{*}(s)$ is always less than the (real) temperature
of the fluid. On the other hand, the effective temperature associated to the
"gas of clusters", $T^{cm}$ is greater than the (real) temperature of the
fluid. In figure \ref{fig5}, we illustrate this inequality which was
already mentioned \cite{SOT98}. In brief, the system can
be seen as a {\it "hot" gas of "cold" physical clusters}.

\begin{figure}[h!] 
\begin{minipage}[t]{40mm}
\includegraphics[scale=.25,angle=0]{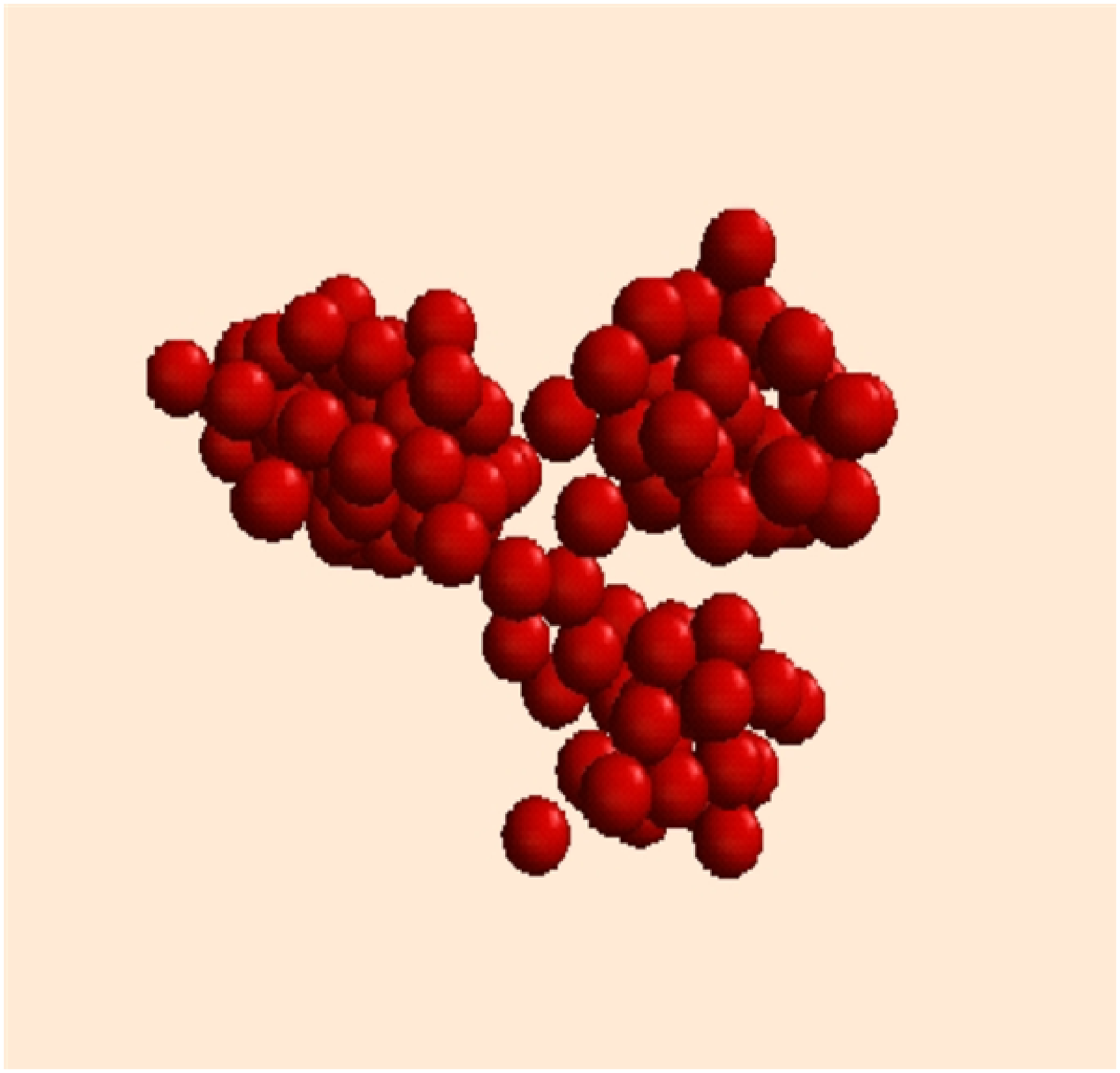}
\begin{center}
$t=0$
\end{center}
\end{minipage}
\hspace{3cm}
\begin{minipage}[t]{40mm}
\includegraphics[scale=.25,angle=0]{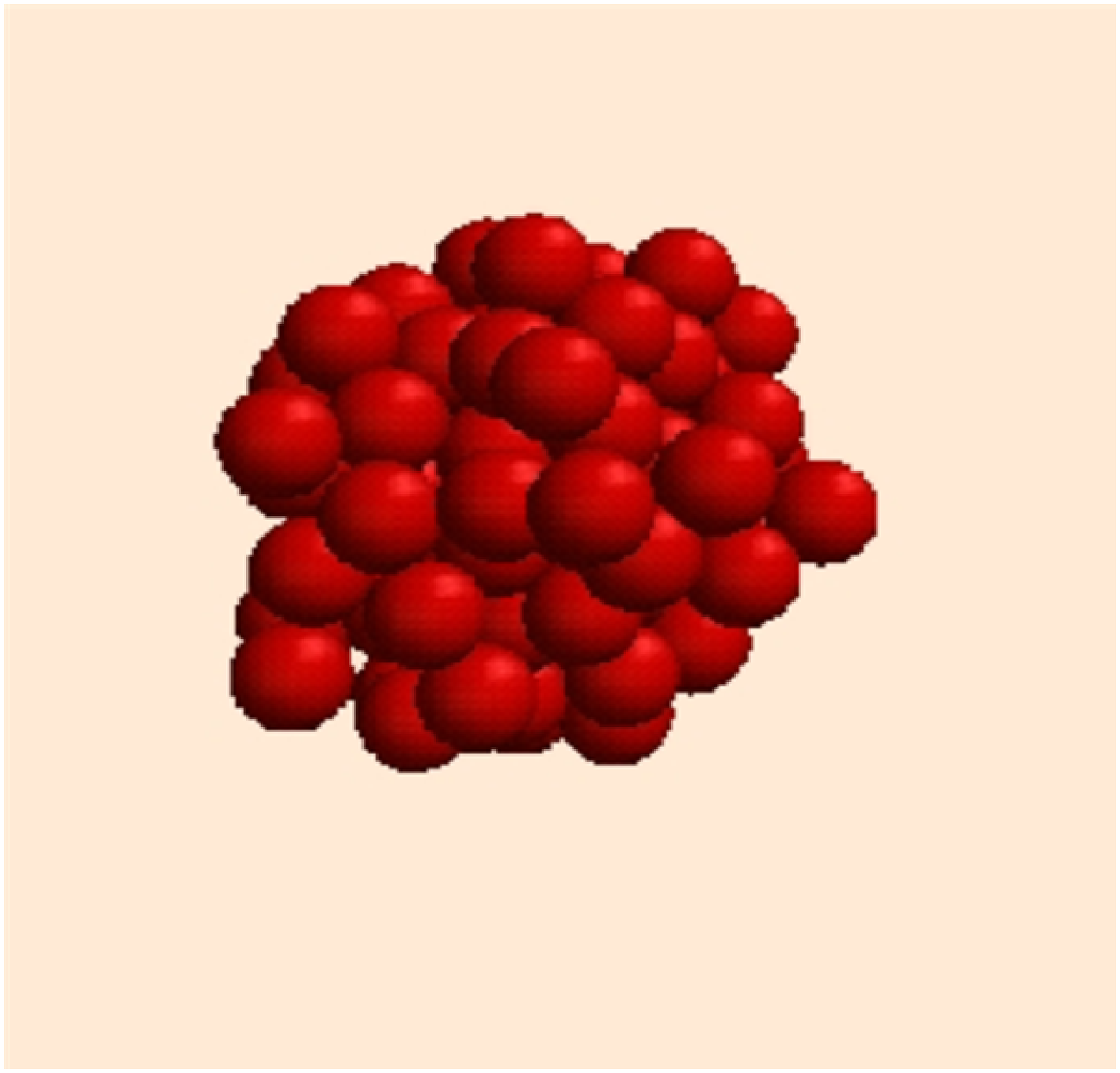}
\begin{center}
$t=400\tau_0$
\end{center}
\end{minipage}
\caption{\label{fig89} \it \footnotesize Typical shape of the largest
cluster in the confined medium ($t=0$) and at asymptotic times after expansion
($t=400\tau_0$).}
\end{figure}

The present results show that various basic hypothesis of the SEM
\cite{SMM,GSI_Ns} are not fulfilled. These models consider an ensemble of
spherical, non-interacting clusters, with an internal temperature equal to
the one of the global system, confined in a volume at low densities
($\rho\sim 0.2$). On the contrary, we find that:
\begin{enumerate}
\item The clusters exhibit fractal shapes in the confined system\footnote{As it
should be, in large systems ($N=12000$), the fractal dimension ($D_f=2.55$)
of the self-bound clusters along the percolation line has been checked to be
in agreement with the one of random percolation \cite{sator1}.} (see figure
\ref{fig89}, left). We must point out that it is not incompatible with the
cold spherical fragments we expect at the end of the expansion. Indeed, by
following the evolution of a self-bound cluster during the expansion (cf
section 3) we find that it becomes spherical as it should be (see figure
\ref{fig89}, right).
\item Even at densities as low as $\rho=0.1-0.2$, the strong
  interaction (without Coulomb) between clusters still represents
  about 40 \% of the total potential energy.
\item The internal temperature of the clusters is lower than the temperature of
   the system. Thus, this fragments will become in the course of the expansion
   (see next section) the observed fragments, without significant particle
    evaporation.

\end{enumerate}

Usually three observables are used to extract the temperature from nuclear
experimental data: i) the kinetic energies of the fragments, ii) the
population of the excited states of fragments \cite{B_Tsang} and iii) the
ratio of isotopic yields \cite{albergo}. All these methods assume that these
observables will yield the thermodynamic temperature of the system, $T$. Our
study contradicts this. As shown in figure 5, the internal and ``kinetic
energy'' temperatures of the fragments are not a measure of $T$.

\section{Fragments in Expanding Systems}

Having studied the properties of clusters in confined systems, we now
turn to the problem of their evolution once the system is allowed to
expand freely. The questions we want to address are :
\begin{itemize}
\item How and when do the fragments appear ? Does the asymptotic fragment
size distribution and kinetic energies relate to the initial configuration
or to an intermediate freeze-out configuration~?
\item Does the system follow a quasi-static path up to some, low
  density, freeze-out configuration where the final fragment
  distribution would be determined, as assumed in the standard
  \cite{MMC,SMM,Friedman,GSI_Ns} models~?
\item How can the temperature and density of this initial configuration be
  inferred from the information measured at asymptotic times ?
\end{itemize}

\begin{figure}[h!] 
\begin{center}
\includegraphics[width=7.5cm,angle=-0]{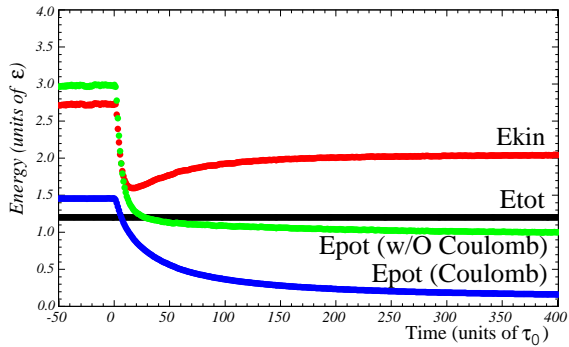}
\end{center}
\caption{\label{fig10} \it \footnotesize Energies as a function of
 time for a CMD calculation performed for $E_{tot}=1.2$, $N=189$
 and $\rho=0.8$. Are shown the total energy, the total kinetic energy,
 the potential energy (absolute value, not including Coulomb) and the
 Coulomb potential energy. The container is removed at time t=0 (see
 text).}
\end{figure}

In order to answer these questions, we have analyzed a number of
expansions of the system using the CMD with
Lennard-Jones plus Coulomb potentials \cite{CMD1}. The system is
characterized, as in section 2, by its energy (microcanonical ensemble), its
density and the number of particles.  
It is enclosed in a cubic
container with perfectly reflective walls. The
calculations proceed in two steps~:
\begin{itemize}
\item After a thermalisation period, the container is removed at a time
defined as $t=0$.

\item The system is then allowed to expand freely during a period of
  time sufficient to establish the final (asymptotic) fragment size
  distribution.
\end{itemize}
Let us first examine the time evolution of the components of the energy.
Figure \ref{fig10} displays, the evolution of the L-J
potential energy, the Coulomb energy, the total kinetic energy and the total
energy.
At $t=0$, the effect of the opening of the container is clearly visible by a
rapid decrease of the potential energy. This corresponds to the fact that
the mean distance between clusters increases rapidly. It can be noted that
the asymptotic potential energy does not go to zero, indicating the
existence of fragments of finite size. The kinetic energy shows a complementary
behaviour : after a sharp decrease, the Coulomb acceleration is clearly
observed.
\begin{figure}[h!] 
\begin{center}
\includegraphics[width=7.5cm,angle=-0]{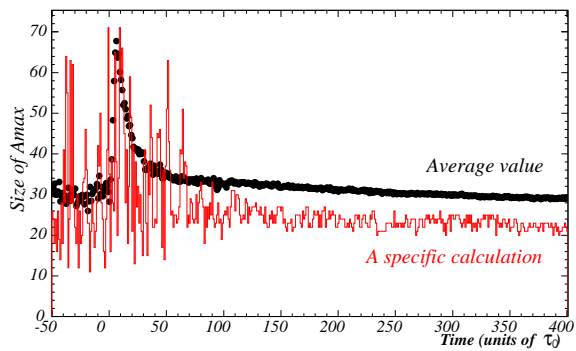}
\end{center}
\caption{\label{fig12} \it \footnotesize Time evolution of the size of the
largest fragment. Full dots correspond to the average values for a
100 calculations, the line to a specific calculation. The container is
removed at time $t=0$ (see text). CMD calculations with $E_{tot}=1.2$, $N=189$
and $\rho=0.8$.}
\end{figure}

We shall now study the evolution of the size of the largest
fragment (Figure \ref{fig12}). The line gives the result of a single event
whereas the full dots represent an average over 100 expansions. The striking
feature of this figure is the observation that the average value of the
largest \emph{fragment} at asymptotic times is very close to the average
value of the largest \emph{cluster} identified in the container ($t<0$). The
correspondence between the largest cluster in the container (time average)
and the asymptotic largest fragment (event average) is systematically
observed. This conclusion is valid for all the calculations that we have
performed, independently of the initial density and energy.

\begin{figure}[h!] 
\begin{center}
\includegraphics[width=11cm,angle=-0]{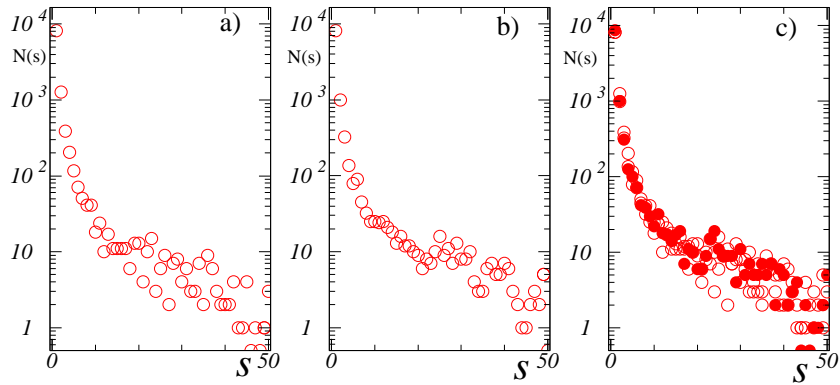}
\end{center}
\caption{\label{fig13} \it \footnotesize Fragment size distribution averaged
over $100$ events for $t=0$, $200\tau_0$ and $400\tau_0$. CMD calculations
with $E_{tot}=1.2$, $N=189$ and $\rho=0.8$. The figure $a)$ corresponds to
$t=0$, $b)$ to $t=200\tau_0$ and $c)$ to $t=400\tau_0$. Full circles
indicate the superimposed results of $a)$ and $b)$.}
\end{figure}

\begin{figure}[h!] 
\begin{center}
\includegraphics[width=7.5cm,angle=-0]{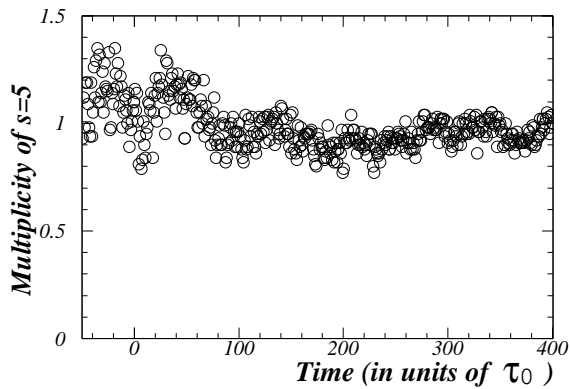}
\end{center}
\caption{\label{fig14} \it \footnotesize Average multiplicity of
fragments of size $s=5$ as a function of time. The container is removed at
time $t=0$ (see text). CMD calculations with $E_{tot}=1.2$, $N=189$ and
$\rho=0.8$.}
\end{figure}
Figure \ref{fig13} shows confirmation of this conclusion by the examination
of more detailed quantities. It presents the fragment size distributions as
a function of time. To within
statistical fluctuations, one observes that they are almost identical.
Figure \ref{fig14} shows the evolution in time of the multiplicity of
fragments of size $s=5$. There again, the asymptotic value is very close to
the initial one. In view of the rapid decrease of the fragment size
distribution observed (figure \ref{fig13}), the stability of this quantity
is impressive.

The correlation between the initial and final size distributions is,
we believe, a consequence of the violent expansion phase (which
follows the opening of the container), the small size of the
system and the presence of the Coulomb repulsion between the clusters.
Because of the small size of the system, the fragments are always
close to the surface and will escape freely into the vacuum. For a very
large system, the expansion phase would give rise, in the central
region, to substantial thermal and chemical activity and hence modify
the thermodynamic parameters of the system. In a small system, this
modification appears almost unobservable.
The influence of the expansion phase is enhanced by the presence of
the Coulomb force. The Coulomb potential not only accelerates the
expansion but also introduces, between the fragments, a Coulomb
barrier, which inhibits particle exchanges.
\begin{figure}[h!] 
\begin{center}
\includegraphics[scale=0.4,angle=-0]{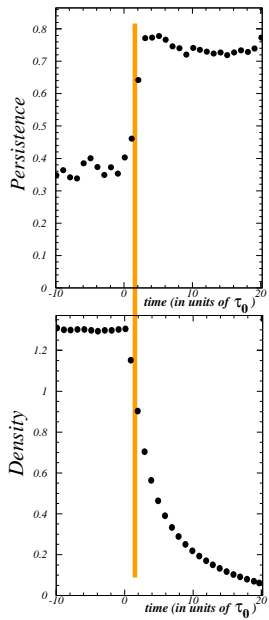}
\end{center}
\caption{\label{fig14bis} \it \footnotesize Persistence (see text) of
fragments of size $s>3$ and density of the expanding system, as a function
of time. $E_{tot}=1.2$, $N=189$ and $\rho=1.3$.}
\end{figure}

The violence of the expansion phase is illustrated by the figure
\ref{fig14bis} which shows the \emph{persistence} of fragments and the
density of the system, as a function of time. The persistence is defined as
the fraction of the number of particles present in a given fragment which
will remain in this fragment at asymptotic times ($t=400\tau_0$). The
density, a poorly defined quantity in such an inhomogeneous expansions, is
deduced from the r.m.s. radius of an uniformly charged sphere of the same
Coulomb energy. Figure \ref{fig14bis} shows that this persistence evolves
very rapidly and almost reaches its asymptotic value at densities of the
order of $0.8$: An exchange of particles between fragments is clearly
observed during this (short) period, but this limited chemical activity does
not induce a modification of the average size of the fragments.

The picture that comes out from these calculations therefore corresponds to
a fast, non equilibrium, expansion of a gas of clusters. This expansion acts
as a "developer" of the presence of these clusters, the Coulomb repulsion
helping to "fix" them. Thus, the (event) average distribution of fragments
reflects the (time) average distribution of clusters in the dense and hot
initial state.

\begin{figure}

  \begin{center}
\includegraphics[scale=0.4]{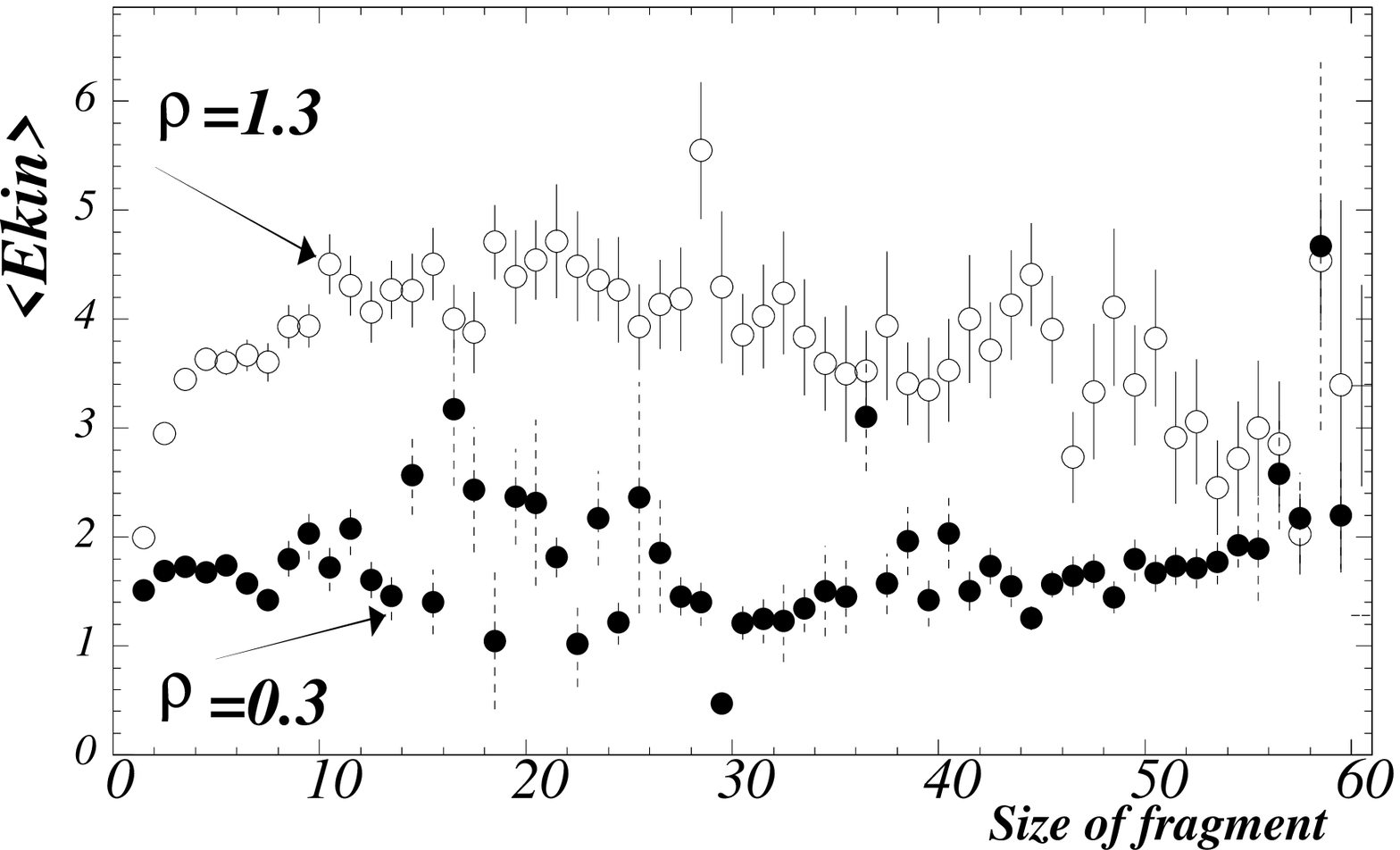}
  \end{center}

\caption{\label{fig15} \it \footnotesize Distributions of the kinetic
energies as a function of the fragment size for a CMD calculation with
$E_{tot}=0.2$, $N=189$ and  densities of 1.3 (empty dots) and
0.3 (full dots). Calculations does not include Coulomb interaction.}

         \begin{center}
         \includegraphics[scale=0.4,angle=-0]{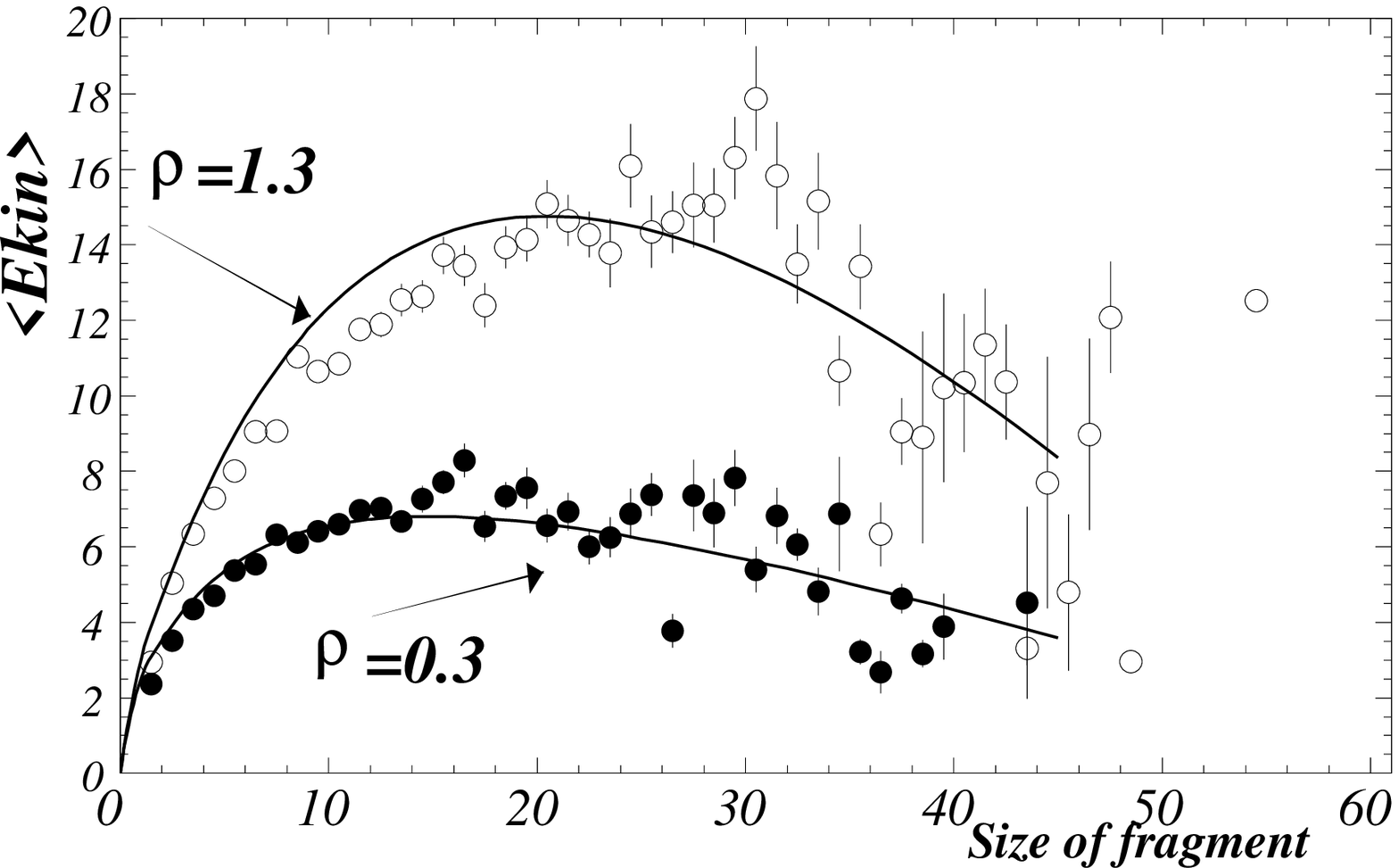}
         \end{center}

 \caption{\label{fig16}\it \footnotesize Same as figure
\ref{fig15}, but with Coulomb interaction. Remark the difference in
the $E_{kin}$ scales. See text for the meaning of the solid line.}

\end{figure}

We now study the kinetic energies of the fragments. Figure \ref{fig15} shows
the asymptotic kinetic energy of expansions \emph{without} Coulomb
interaction ($E_{tot}=0.2$). As can be seen, the kinetic energies obtained
from different starting densities (0.3 and 1.3) are quantitatively
different. This demonstrates once more that the expansion is a
non-equilibrium phenomenon. If equilibrium, at a given energy, was
maintained during the expansion, the system starting from $\rho=1.3$ would,
at a point in time, reach the same point in $\{T-\rho\}$ space as the one
from which the system at $\rho=0.3$ starts and hence the two systems would
be indistinguishable in all respect: Figure \ref{fig15} demonstrates that it
is not the case.

Figure \ref{fig16} shows the kinetic energies in the case
\emph{with} Coulomb interaction. The shape of these distributions is
characteristic of the presence of Coulomb forces and is close
to what is observed in nuclear reactions at Fermi bombarding energies
($\sim 50$ MeV/A) for symmetric nuclei \cite{Ekins1,Ekins3}.
The solid lines show a qualitative
(analytical) prediction of what would be expected in the case of a
uniformly charged medium.

The important feature that is revealed by this figure is the dependence of
the kinetic energies on the initial density. This result appears to confirm
the evolution suggested by the analysis of the fragment size
distribution. Very rapidly after the removal of the container, the fragments
cease to strongly interact with the medium and are accelerated by the
(Coulomb) repulsive potential. Because this expansion takes place out of
equilibrium, the Coulomb energy does not transform into thermal energy but in
fragment kinetic energy. The higher the
density of the initial configuration, the greater the final kinetic
energies are. We can therefore consider that the asymptotic kinetic energies are a
``measure'' of the density at which the systems departs from
equilibrium.

This scenario accounts, at least partially, for the radial flow often
associated with multifragmentation processes \cite{Ekins1,Ekins3}. As stated
above (section 1), the standard method of analysis, using the SEM, assumes a
low density configuration and therefore naturally underpredicts the kinetic
energies of the fragments when the system has evolved (out of equilibrium)
from a high density state. To compensate for this, an extra component of
``radial flow'' is artificially introduced. In the scenario we suggest, such
an extra flow is not necessary when the calculation is performed at the
correct initial density. It is important to remark that our scenario does
not necessarily imply large values for the kinetic energy. For example, for
peripheral heavy ion reactions, no initial compression is expected. Even
more, for proton-nucleus or $\pi$-nucleus collisions, initial densities
lower than the normal equilibrium nuclear density are plausible
because of the formation of holes inside a nucleus with normal volume
\cite{GA91}. 

Having ``measured'' the density, we now turn to the estimation of the
temperature. The sensitivity of the kinetic energies to the density,
provides a ``method'' to determine the thermodynamic parameters of the point
at which the system departs from equilibrium. The analysis of the fragment
size distribution will select which iso-$n(s)$ line is to be considered. The
analysis of the kinetic energies will determine the density. The
intersection of both will yield $T$ and $\rho$. Evidently, this method is
physics (model) dependent, i.e. it is only if one is able to calculate the
relationship $E(T,\rho)$ and the fragment size distribution at each
$(T,\rho)$ point that the thermodynamical parameters can be extracted from
the analysis of the data measured at asymptotic times.

\section{Final remarks}

We have explored a new scenario of multifragmentation, based on the
observation that a dense and hot fluid, at equilibrium, can be viewed
as a hot gas of cold clusters. These clusters are defined as
self-bound ensembles of particles. When the system is allowed to
expand freely, it proceeds, \emph{out of equilibrium}, as an ensemble of
interacting clusters. Once these
clusters cease to interact with  each other, they become the
observable fragments.

We believe that this "Little Big Bang" scenario is not only more realistic
than the "standard" low density freeze-out model but that it also allows
to solve a number of recurrent problems: 
\begin{itemize}
\item Equilibrium is assumed only at
the beginning of the process, when the high density and temperature make
this assumption more likely. No "quasi-static" hypothesis of expansion up to
a low freeze-out density is needed. 
\item Due to this compacity, the Coulomb repulsion between fragments
generates larger kinetic energies. Clusters in the dense medium have
"ramified" (fractal in large systems) forms, allowing more compact
configurations in the initial stage. The observed radial flows can thus be
explained, at least partially, in a natural way.
\item Isolated fragments are cold, because the precursor
clusters are already cooler than the ensemble of the system. 
\end{itemize}

The aim of this work was to present a new scenario of
multifragmentation, not a well finished model producing results
directly comparable with experimental data. One has to keep in mind that the
present scenario is supported by classical molecular dynamics
simulations of the expansion of a system of particles interacting
through Lennard-Jones plus Coulomb potentials, initially confined in
a container. Although we believe that these results are 
generic for describing the free expansion of any simple fluid (and
that nuclear matter at high temperatures behaves as a simple fluid),
it is necessary to confront our results with other
calculations.  Namely, with Quantum Molecular Dynamics \cite{QMD1,QMD2}
 calculations to test the equilibrium hypothesis in the
first stages of the reaction, and with Fermionic Molecular Dynamics
\cite{FMD} to check the importance of quantum effects,
particularly at sub-critical temperatures.

We would like to thank F. Lavaud for his contribution to the expansion
phase calculations.  One of us (N.S.) wishes to acknowledge financial
support of the European TMR Network-Fractals (Contract number:
FMRXCT-980183).

\end{document}